\documentclass[
    ,final            % use final for the camera ready runs
  ,numberedheadings % uncomment this option for numbered sections
  ]
  {aipproc}

\layoutstyle{6x9}

\def\varv{\upsilon}

\begin{document}

\title{Spectroscopy and hydrodynamics of dense stellar winds}



\author{Wolf-Rainer Hamann}{address={Universit\"at Potsdam, Germany}}

\author{G\"otz Gr\"afener}{address={Universit\"at Potsdam, Germany},
  ,altaddress={Armagh Observatory, Northern Ireland}} 

\author{Lidia M. Oskinova}{address={Universit\"at Potsdam, Germany}}
\author{Achim Feldmeier}{address={Universit\"at Potsdam, Germany}}

\begin{abstract}
Analyzing the spectra from Wolf-Rayet stars requires adequate non-LTE
modeling of their expanding atmosphere. The numerical schemes for
solving the radiative transfer in the co-moving frame of reference have
been developed by Mihalas and co-workers 30 years ago. The most
elaborate codes can cope today with many hundred explicit non-LTE levels
or super-levels and account for metal-line blanketing.

The limited agreement with observed spectra indicates that the model
simplifications are still severe. One approximation that has to be
blamed is homogeneity. Stellar-wind clumping on small scales was
easily implemented, while "macro-clumping" is still a big challenge. First
studies showed that macro-clumping can reduce the strength of predicted
P-Cygni line profiles in O-star spectra, and largely affects the X-ray
line spectra from stellar winds.

The classical model for radiation-driven winds by Castor, Abbot and
Klein fails to explain the very dense winds from Wolf-Rayet stars. Only
when we solved the detailed non-LTE radiative transfer consistently with the
hydrodynamic equations, mass-loss rates above the single-scattering
limit have been obtained.  
\end{abstract}

\maketitle

%%%%%%%%%%%%%%%%%%%%%%%%%%%%%%%%%%%%%%%%%%%%
%% MAINMATTER
%%%%%%%%%%%%%%%%%%%%%%%%%%%%%%%%%%%%%%%%%%%%

\section{Modeling the spectra of dense stellar winds}

Dimitri Mihalas and co-workers developed the formalism and the numerical 
algorithms to model the atmospheres of hot stars in non-LTE. After
having accomplished this task for the static, plane-parallel case,
the problem of spherically expanding atmospheres was attacked in a
series of seminal papers by Mihalas, Kunasz, and Hummer in the years
1975/76. 

One important step was to realize that the radiative transfer in highly 
supersonic flows, where  the Doppler shifts are much larger than the
width of the line absorption profile, are treated most conveniently in
the co-moving frame of reference (CMF). For the equations of statistical
equilibrium the ``radiative transition rates'' are required, and these
have to be evaluated with the (angle-averaged) radiation intensity in
the CMF as it is ``seen'' by the matter.
As a drawback, the equation of radiative transfer becomes a {\em
partial} differential equation in the CMF: because the frequency $\nu$
is measured with respect to the co-moving frame, a photon is
changing its frequency when propagating through the differentially
moving medium. For a ray with coordinate $z$, a velocity $\varv$ in radial
direction, and angle $\vartheta = \arccos \mu$ between the
radial direction and the ray, the CMF transfer is 

\begin{equation}
\frac{\partial I_\nu}{\partial z} - 
\frac{\nu}{c}\frac{{\rm d} (\mu \varv)}{{\rm d}z}~ 
\frac{\partial I_\nu}{\partial \nu} 
= \kappa\ (I_\nu - S_\nu)\ .
\label{eq:rt}
\end{equation}

In spherically extended atmospheres, the radiation field becomes very
anisotropic, and hence many rays of different angle (or, impact
parameter) are needed for an accurate numerical representation. However,
for isotropic opacity the radiative rates depend only 
on the angle-averaged radiation intensity $J_\nu$. Therefore it is
sufficient, and much more economic, to solve the {\em moment} equations 
of the radiative transfer. Most of the iterations can be performed by
solving the moment equations with given Eddington factors. Only from time to
time the latter must be updated. For this purpose the moments of the
radiation field must be newly integrated (Eq.\,(\ref{eq:momdef})) from
the angle-dependent intensity, which is obtained from solving the
angle-dependent transfer equation for many rays.

As a closure to the moment equations, Auer \& Mihalas (1970) introduced
the ``method of variable Eddington factors'' for the static case. In
Mihalas, Kunasz \& Hummer (1976) they extended this concept to
spherically expanding atmospheres. {\em Four} moments of the radiation
intensity are needed in this case:
\begin{equation}
[J_\nu, H_\nu, K_\nu, N_\nu] = \int {\rm d}\mu\ I_\nu~ 
[1, \mu,\mu^2, \mu^3]\ .
\label{eq:momdef} 
\end{equation}
Eddington factors are defined as $f = K_\nu / J_\nu$ (like in the static
case) and $g = N_\nu / H_\nu$. 

In the numerical application of this method, we encountered problems
with both of these definitions. On physical grounds, the intensity-like
moment $J_\nu$ should be positive. Solving the angle-dependent CMF
transfer Eq.\,(\ref{eq:rt}) by means of a Feautrier-like differencing
scheme, however, sometimes yields slightly negative values at some
frequency and radius points, just due to the limited numerical accuracy.
This can make the definition of the Eddington factor $f = K_\nu / J_\nu$
singular. We overcame this problem by solving the angle-dependent
radiative transfer with an {\em integration} method along short
characteristics. Such method can guarantee positive intensities.

The problem with the other Eddington factor, $g = N_\nu / H_\nu$, is
even more severe. There is actually no reason why the Eddington flux
$H_\nu$ may not become negative in some situations. Therefore we replaced the
originally proposed definition of $g$ by $g' = N_\nu / J_\nu$. 
Although now relating
a flux-like to an intensity-like moment, this Eddington factor 
behaves well in the iteration process. 

For the simultaneous solution of the (discretized) radiative transfer
equation and the constraint equations, Mihalas and co-workers composed a
large system of  algebraic equations and solved them by linearization
and elimination techniques. Such method is very much restricted to only
a few atomic levels and transitions. 

\begin{figure}
\includegraphics[width=\textwidth]{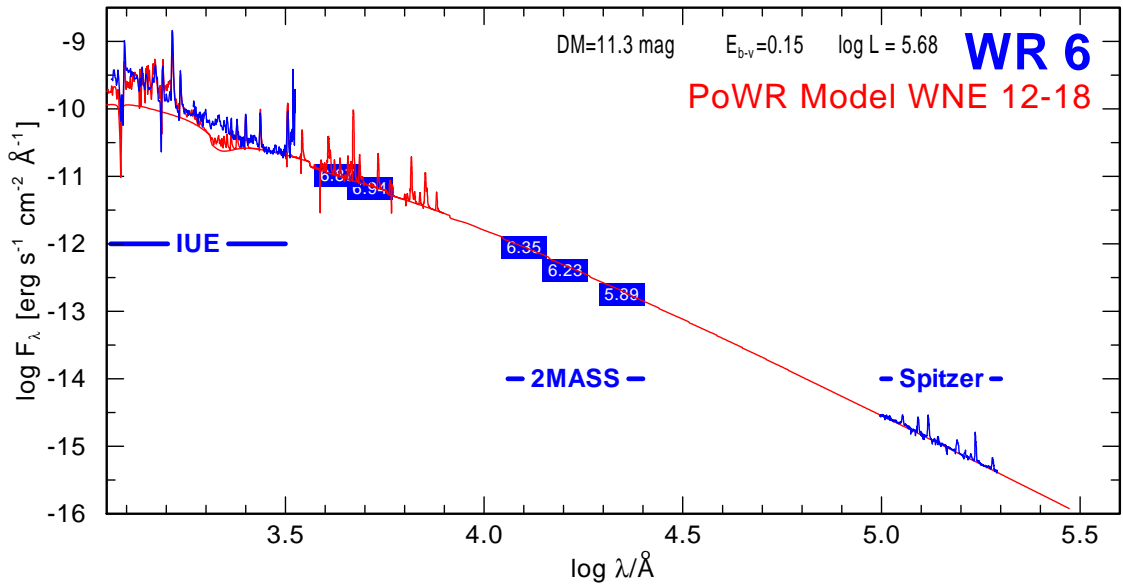}
\caption{Observed (blue) spectral energy distribution of the Wolf-Rayet 
star WR\,6 (WN4) from the UV (IUE) to the mid-IR (Spitzer), compared
to a Potsdam Wolf-Rayet model (red) from the published grid 
({\tt http://www.astro.physik.uni-potsdam.de/PoWR.html})
after applying interstellar reddening.}
\label{fig:WR6}
\end{figure}

Present-day non-LTE codes instead simply solve the radiative transfer
and the statistical equilibrium in turn. Such an iteration scheme, a
so-called lambda iteration, does not converge when large optical depths
are involved. However, in the 1980s it has been discovered that
convergence can be achieved when only the local feedback of the
radiative transfer is consistently incorporated into the statistical
equations. This method, termed ``accelerated lambda iteration (ALI)'' or 
``approximate lambda operator (ALO)'' technique, allows to calculate
model atmospheres with many hundred non-LTE levels, thousands of
line transitions, and even millions of further lines in an approximate 
way as needed for ``iron line blanketing'' (see Fig.\,\ref{fig:WR6} for an
example).

%-----------------------------------------------------
%-----------------------------------------------------

\section{Stellar winds hydrodynamics}

\subsection{Single scattering}

\begin{figure}
\includegraphics[width=0.5\textwidth]{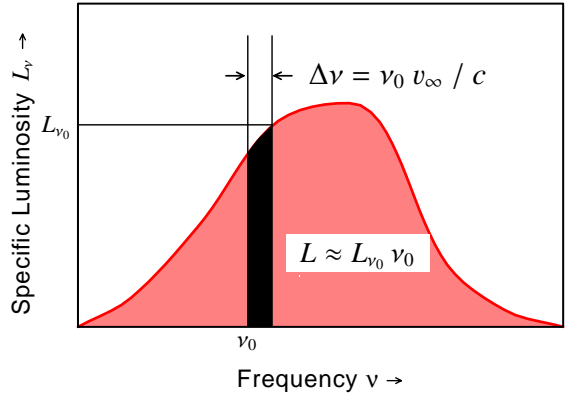}
\caption{Absorption by one optically thick
wind line at about the maximum of the photospheric flux}
\label{fig:linedriving}
\end{figure}

Mass-loss from O stars is driven by the momentum transfer from UV
photons to the ions of heavy elements. This has been worked out in
hydrodynamically self-consistent models for O stars, based on the
fundamental concept of Castor, Abbott \& Klein (1975). The photospheric
radiation is absorbed in a large number of lines, some of them being
optically thick. A line which remains optically thick throughout the wind
sweeps up all photospheric flux over a bandwidth $\Delta\nu = \nu_0\
\varv_\infty / c$, as illustrated in Fig.\,\ref{fig:linedriving}. The
intercepted momentum is then $L_{\nu_0}\ \Delta\nu/c = L\ \varv_\infty/c^2$
where $L_{\nu_0} = L / \nu_0$ is the specific luminosity at the line
frequency $\nu_0$. Comparison with the wind momentum per time, 
$\dot M\ \varv_\infty$, reveals that the mass-loss rate driven by a single
optically thick line is $L/c^2$ -- remarkably just the same rate by
which mass is converted to energy in the stellar interior for sustaining the
luminosity. The actual mass-loss rate in a star can be conveniently
expressed in terms of a $n_{\rm eff}$, the effective number of lines:
$\dot{M} = n_{\rm eff} L/c^2 $. As hydrogen burning consumes about 120
times more fuel than the mass that is coverted into energy, wind mass
loss becomes a signifant driver of stellar evolution when $n_{\rm eff}$
is comparable to this factor. 

Within this model approach, the maximum mass-loss rate is obviously
reached when the whole spectrum is covered by lines and each
photospheric photon delivers its momentum once to the wind, i.e.\ when
$n_{\rm eff} = c/\varv_\infty$  or $\dot{M}=L/(\varv_\infty c)$. 

The mass-loss rates from O and B stars fall below this
``single-scattering limit'', albeit for supergiants like $\zeta$\,Pup not
by much. For many of the Wolf-Rayet stars, however, empirical mass-loss
rates exceed this limit considerably. The CAK models fall
short to explain such dense winds.

%-----------------------------------------------------
%-----------------------------------------------------

\subsection{Multiple scattering}
\label{sect:multiplescattering}

\begin{figure}
\includegraphics[width=\textwidth]{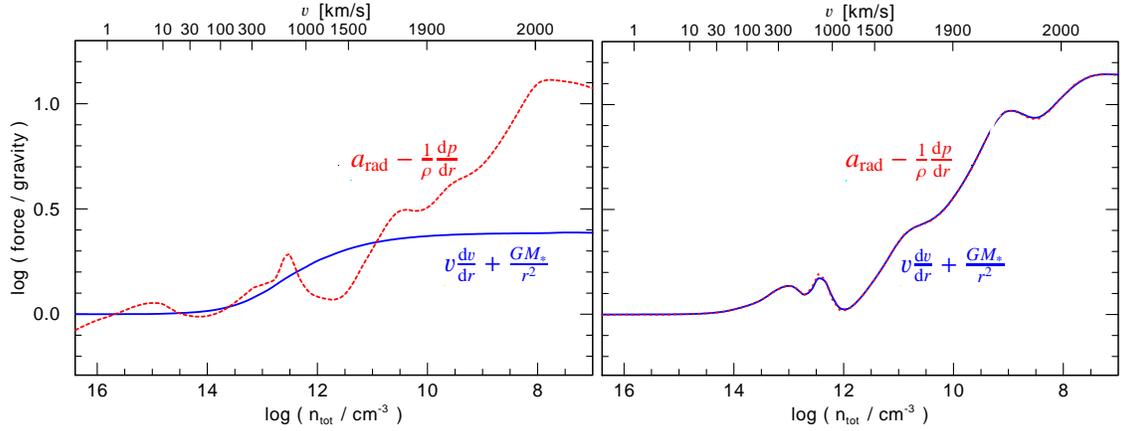}
\caption{Acceleration as function of radius (represented by the atomic
number density) for a WR wind (model for the WC5 star WR\,111, after
Gr\"afener \& Hamann 2005). In order to satisfy the hydrodynamic
equation, the inward forces (inertia plus gravity, solid/blue line) must
be balanced by the outward forces (radiative acceleration $a_{\rm rad}$
and pressure gradient, red/dotted curve). For a model with prescribed
velocity field (``beta law'') and mass-loss rate, the equality of both
terms is not given (left panel). By incorporating the hydrodynamical
equations into the PoWR model atmosphere code, a solution for $\dot{M}$
and $\varv(r)$ is found that is hydrodynamically consistent everywhere in
the wind (right panel).}
\label{fig:combined-acc}
\end{figure}

Mass-loss exceeding the single-scattering limit 
requires that photons deliver their momentum more than
once. Such multiple-scattering effects are not included in CAK-type
models. To overcome this shortfall, we combined the Potsdam Wolf-Rayet
{\sc PoWR} model atmosphere code with the hydrodynamic equations. As the
whole radiative transfer is treated consistently, all possible
multi-line effects are automatically included. Indeed we obtained
self-consistent WR models, as has been demonstrated in Gr\"afener \&
Hamann (2005) for the WC5 prototype WR\,111. 
Figure\,\ref{fig:combined-acc} shows for this model how the inward and
outward forces are perfectly balanced in the consistent, stationary
hydrodynamic solution. Hence we are sure now that the winds of WR stars
can be driven by radiation pressure. The previous shortfalls in
explaining WR winds were obviously due to deficiencies in the
conventional radiative-transfer treatment in CAK-type models.

\begin{figure}
\includegraphics[width=.7\textwidth]{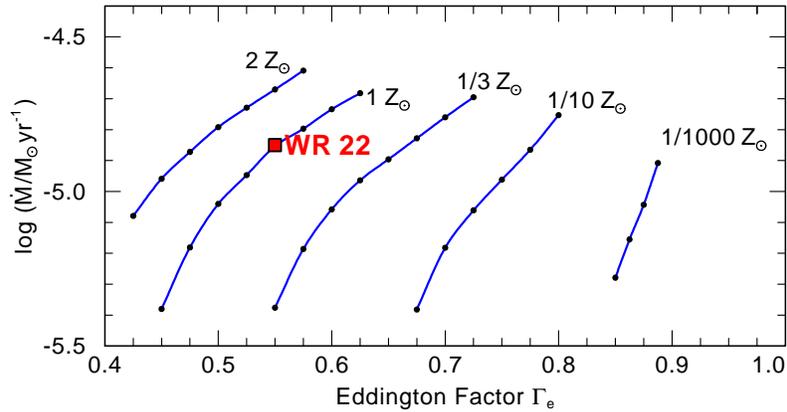}
\caption{Mass-loss rates for models of the same luminosity
($10^{6.3}\,L_\odot$), but different $L/M$ ratio (Eddington factor
$\Gamma$) and different metallicities $Z$ (compared to the solar
metallicity $Z_\odot$). Hydrodynamically consistent PoWR models are
represented by a black dot, blue lines connect models with the same
metallicity.  The empirical parameters of the WN7 star WR\,22 are
indicated by a red square. (From Gr\"afener et al.\ 2008)}
\label{fig:wnl-z}
\end{figure}

Specifically in the dense winds from WR stars the multiple-scattering
effects from the numerous lines of iron-group elements become important.
But why WR stars develop so much denser winds than O stars? We have
studied this question by means of a model grid for luminous WN stars of
late subtype (WNL) in Gr\"afener \& Hamann (2008). According to our
self-consistent HD models, the key point is the $L/M$ ratio (see
Fig.\,\ref{fig:wnl-z}). WR stars are just very close to the Eddington
limit where the radiation pressure on free electrons balances gravity
($\Gamma_{\rm e} = 1$). Mass-loss rates of about $10^{-5} M_\odot/{\rm
yr}$, as typical for WR stars, are obtained for an Eddington
$\Gamma_{\rm e}$ of 0.5 if the metallicity is solar. But even at very
low metallicity, strong winds are predicted if $\Gamma_{\rm e}$ is
sufficiently close to unity.

%-----------------------------------------------------
%-----------------------------------------------------

%-----------------------------------------------------
%-----------------------------------------------------

\subsection{Time-dependent simulations}
\label{sect:hydroshocks}

The hydrodynamic models described in the previous two subsections are 
stationary solutions of the problem. However, these solutions are 
expected to be unstable because of the de-shadowing effect: if one 
volume element would experience a positive perturbation of its velocity, 
it intercepts more radiation in optically thick lines 
and is therefore even more accelerated. 

\begin{figure}
\includegraphics[width=\textwidth]{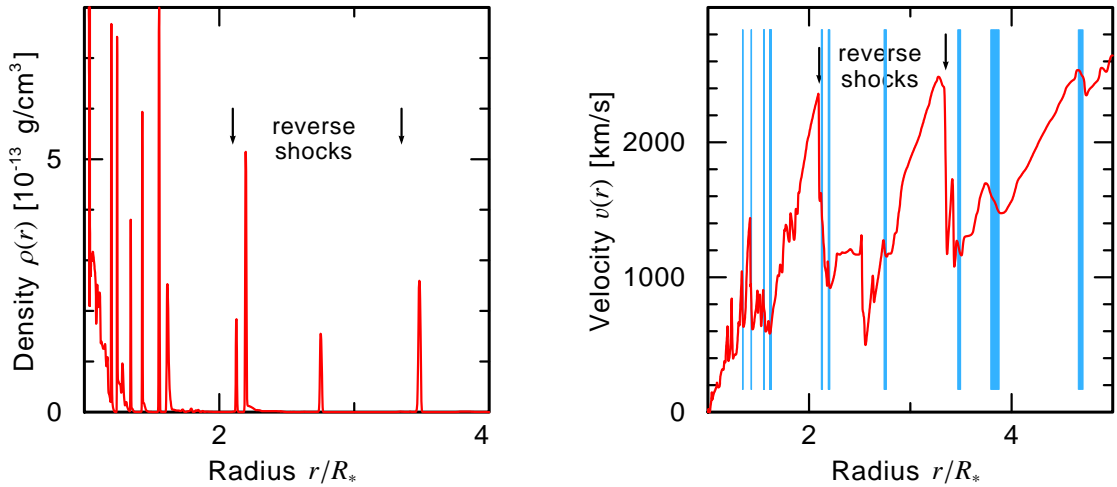}
\caption{Snaphot from a 1D-hydrodynamic simulation of a line-driven 
stellar wind (adapted from Feldmeier et al.\ 1997). The strong peaks in
density (left panel) contain most of the matter. This motivates the {\em
fragmented shell} model of our X-ray studies. The radially compressed
shells are created by reverse shocks characterized by strong velocity jumps
(right panel). The shaded stripes in the velocity plot indicate the
location and radial extension of the overdense shells.} 
\label{fig:shocks}
\end{figure}

The effect of this instability can be seen in the time-dependent 
hydrodynamical simulations presented by Feldmeier et al.\ (1997). These 
calculations are for spherical symmetry, but account for heating and 
cooling processes. A snapshot for an O star wind is shown in 
Fig.\,\ref{fig:shocks}. The instability leads to the formation of dense 
shells, which contain almost all of the wind material and have a radial 
separation of about one stellar radius. These structures are created by 
strong reverse shocks with velocity jumps of several hundred km/s. The 
shocked gas reaches temperatures of a few million Kelvin.  

Corresponding multidimensional simulations which account for the
radiative force on lines in non-radial directions and for the the heating
and cooling of the shocks are not yet available.

%-----------------------------------------------------
%-----------------------------------------------------

\section{Inhomogeneous stellar winds}

\subsection{Microclumping}
\label{sect:microclumping}

As discussed in Sect.\,\ref{sect:hydroshocks}, hot-star winds are
expected to be structured by the hydrodynamic instability that is
inherent to the line-driving mechanism. This prediction is supported by
various observational evidences for wind variability and inhomogeneity.
An important spectroscopic fingerprint of wind ``clumping'' had been
pointed out already by Hillier (1984), namely the electron-scattering
wings of strong emission lines. The ``Thomson scattering'' of photons by
free electrons is often considered as being coherent. However, this is
only correct in the rest frame of the electron (when neglecting the
Compton shift due to the recoil), but not when considering the
electron's thermal motion. In fact, the thermal speed of the electrons
is very high, about 500\,km/s at 10\,kK, and therefore the corresponding
Doppler shift causes a redistribution in frequency that is comparable
with the width of wind-broadened emission lines. 

The emission lines that dominate the spectra of Wolf-Rayet stars are
mainly fed by recombination cascades. The recombination process scales
with the square of the density. The Thomson scattering, in contrast,
scales linearly with the (electron) density. Considering an
inhomogeneous medium, the ratio between the mean of the density-squared
and the square of the mean density, $\langle \rho^2 \rangle / \langle
\rho \rangle^2$, becomes larger than unity. Therefore, the fit of
emission lines and their electron-scattering winds can provide a tool to
estimate the degree of stellar-wind clumping.

In a simple approach, one can assume that the whole stellar wind is
concentrated in clumps, while the interclump medium is void. When the
clumps fill a fraction $f_{\rm V}$ of the volume, the density in the
clumps is by a factor $D = f_{\rm V}^{-1}$ enhanced over a smooth
wind. 

A further simplifying assumption is that the individual clumps are
optically thin at all frequencies. This approximation is termed
``microclumping''. It can be easily implemented  in model atmosphere
codes, as it leaves the radiation field homogeneous. For Wolf-Rayet
stars, the electron-scattering wings of strong emission lines are
usually well reproduced with this formalism when the clumping contrast
$D$ is set to values between 4 and 10 (e.g.\ Hamann \& Koesterke 1998). 
The spectra of O and B stars,  however, do not show strong enough
emission lines for which this method could be applied. There are
empirical indications that clumping develops already deep in the
wind where the expansion velocity is still small, and decreases 
at larger distance from the star (Puls et al.\ 2006).

A major consequence of microclumping is a reduction of empirical mass-loss 
rates that are derived from $\rho^2$-depending processes, such as
emission lines fed by recombination cascades, as well as the free-free
radio emission. Compared to ``unclumped'' diagnostic, this reduction is
by a factor of $\sqrt{D}$. 

In a recent work, Zsarg\'o et al.\ (2008) introduced a second, diluted
wind component that fills the space between the clumps. This thin
interclump gas becomes more easily ionized, especially from hard
radiation produced in wind shocks, and therefore contributes to the
spectral lines from high  ions, e.g., from O\,{\sc vi}.

\subsection{Macroclumping}

The {\em microclumping} assumption that the individual clumps have only
small optical depth cannot be justified for all frequencies. The
wind-compressed shells in the hydrodynamical simulations (cf.\
Fig.\,\ref{fig:shocks}) have a thickness of the order of 0.1\,$R_\ast$.
Estimates from variability lead to similar clump diameters. 
In the winds from WR
and O-type stars, clumps of such size are expected to be optically thick
in the cores of strong lines, and also in the far-UV and soft X-ray
continua from bound-free and K-shell absorption. Therefore we introduce 
a more general concept which accounts for the finite optical thickness 
of clumps. However, drastic approximations are necessary to keep the 
problem manageable.

\subsubsection{Macroclumping and X-rays}

%-----------------------------------------------------
\begin{figure}
\includegraphics[width=\textwidth]{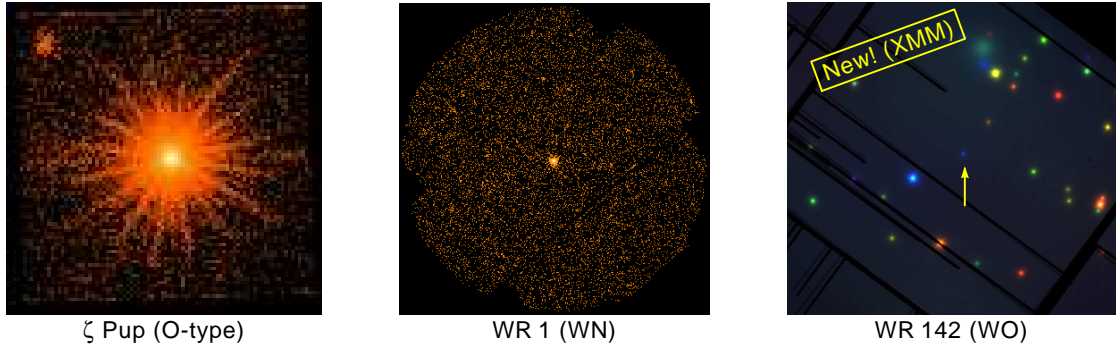}
\caption{Illustration of the different X-ray brightness of stellar winds.
While O-star winds are often strong X-ray sources, only some of the WN
stars show a measurable X-ray flux, when all binaries are disregarded.
As the first, albeit faint X-ray source among the WO/WC class, we
recently detected the very hot and compact WO star WR\,142.}
\label{fig:3xrays}
\end{figure}
%-----------------------------------------------------

Early-type stars can be often observed as X-ray sources. Many such
sources are found to be binaries, where the X-rays may be produced by
the collision of two stellar winds. However, there are also many
putatively single early-type stars with X-ray detections.

%-----------------------------------------------------
\begin{figure}
\includegraphics[width=0.5\textwidth]{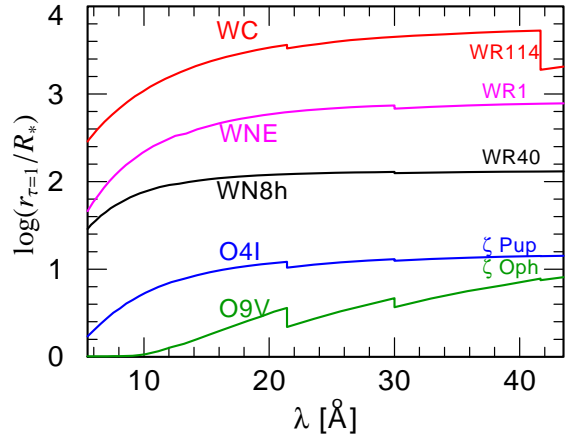}
\caption{Radius where the radial optical depth reaches unity, plotted
versus wavelength in the X-ray range. The high continuum opacity is due
to bound-free and K-shell absorption. The models are for typical
parameters, but without clumping.}
\label{fig:tau1}
\end{figure}
%-----------------------------------------------------

X-ray emission is believed to originate from hot, shocked gas embedded
in the stellar winds. There are two arguments why the X-rays are
production must be located in the inner wind region, firstly because the
line-driving instability should be largest where the acceleration is
strongest, and secondly because the X-ray line profiles are typically
wind-broadened with only about half of the terminal wind velocity. In order
to be observable, this X-ray emission has to escape from the expanding
atmosphere through the absorbing ``cool'' wind component which contains
the bulk of the matter. This wind attenuation may account for the
observed decrease of the emergent X-ray flux with increasing density of
the stellar wind illustrated in Fig.\,\ref{fig:3xrays}. 

Checking this wind attenuation quantitatively reveals a severe problem,
as demonstrated in Fig.\,\ref{fig:tau1}. For X-ray wavelengths where
emission lines are observed, the plot shows the radius where the radial
optical depth reaches unity. According to these predictions from typical
models, X-rays should not be able to escape from lower wind regions, at
least not for an O supergiant like $\zeta$\,Pup, and definitely not for
Wolf-Rayet stars.

An escape from this dilemma can be searched by reducing the stellar
mass-loss rates by adopting very high clumping factors $D$. At least in
the case of WN stars this runs in various contradictions, e.g.\ with the
electron-scattering line wings (see above), or with the mass-loss rate
of the binary V444\,Cyg from its period change.  

Alternatively, whe have shown that the X-ray observations can be
explained when the macroscopic size of the clumps is taken into account.
The X-rays lines are emitted from hot shocks, while the absorption takes
place in the cool wind component by continuum processes. This decoupling
of emission and absorption greatly facilitates a semi-empirical
modeling.

Monte-Carlo calculations have been applied to model the emergent X-ray
line profiles with randomly distributed line emitters and absorbing
clumps (Oskinova et al.\ 2004, 2006). Basically the same results  are
obtained from an analytical treatment that holds in the limit that the
emitting spots and the absorbing clumps are numerous enough for a
statistical description (Feldmeier et al.\ 2003, Owocki \& Cohen 2006).
The consequence of the wind clumping is a reduction of the effective
opacity. Some atomic absorbers are hiding in the shadow of others, while
the gaps between the shell fragments open channels where photons can
travel far.
  
%-----------------------------------------------------
\begin{figure}
\includegraphics[width=\textwidth]{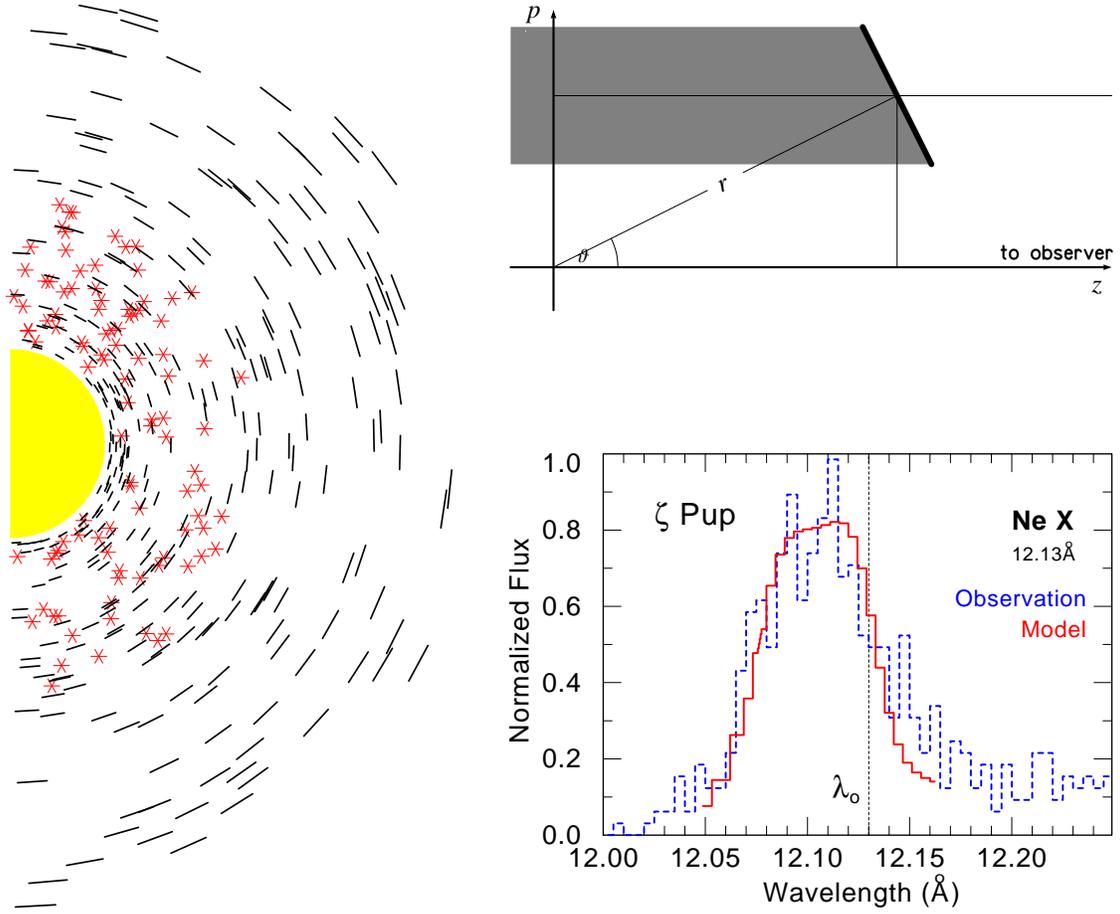}
\caption{The {\em fragmented shell model} for the formation of X-ray
lines. The radiation is emitted from randomly distributed shocks (red
stars in the left sketch) and absorbed by radially compressed shell
fragments. In the radiative transfer calculations, these fragments are
treated as thin slabs, leading to an angle-dependent opacity (see
upper-right sketch). Only with these assumptions the model (red) can
reproduce the shifted and skewed X-ray line profiles as observed
(blue-dashed).}
\label{fig:xmodel}
\end{figure}
%-----------------------------------------------------

The Potsdam work (Feldmeier et al.\ 2003, Oskinova et al.\ 2004, 2006)
found that the observed X-ray line profiles can be explained best if the
``clumps'' are assumed to have the shape of ``pancakes'' or ``shell
fragments'' (see Fig.\,\ref{fig:xmodel}). This idea is also supported
from the theoretical consideration that the forces which compress the
clumps act mainly in radial direction. If clumps are anisotropic, their
{\em projected} cross section leads to a ``venetian blind effect'',
where in the limit of entirely flat fragments the opacity scales
proportional to $\mu = \cos \vartheta$. With the corresponding effective
opacity becoming angle-dependent as $\kappa_{\rm eff} \propto \mu$, the
optical depth increment for a ray of any impact parameter grows with
the change of radius: ${\rm d}\tau \propto |{\rm d}r|$, in contrast to 
the isotropic case where ${\rm d}\tau = \kappa\ {\rm d}z$.

\subsubsection{Macroclumping and line formation}

%-----------------------------------------------------
\begin{figure}
\includegraphics[width=0.7\textwidth]{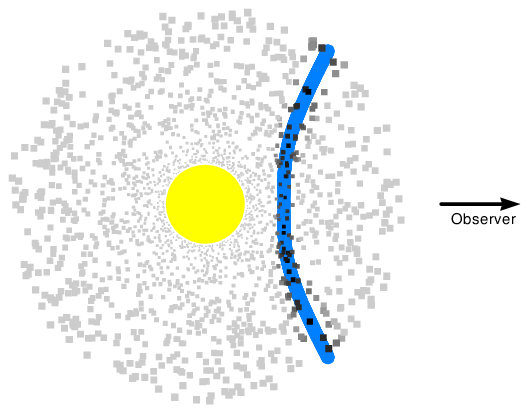}
\caption{Porosity in the case of line opacity. At a given observer's
frame frequency, rays can only interact with those clumps (dark shaded)
that are close to the {\em constant radial velocity surface} (CRVS, blue
thick curve), while other clumps (light grey) are out of the line
resonance.}
\label{fig:clumps}
\end{figure}
%-----------------------------------------------------

In Oskinova et al.\ (2007) we considered specifically the case of a
spectral {\em line} formation. We take the statistical approach as
already used for the X-ray absorption, but in the line case we restrict
ourselves to isotropic clumps which have the same optical diameter
$\tau_{\rm clump}$ in all directions. At a given location, all clumps
have uniform size. Under these assumptions, the {\em effective} opacity
becomes
\begin{equation}
\kappa_{\rm eff} = h^{-1}\ \left( 1 - e^{-\tau_{\rm clump}} \right) 
= \kappa_{\rm smooth}\ \frac{1 - e^{-\tau_{\rm clump}}}{\tau_{\rm clump}}
\end{equation}
where $h$ is the well-known ``porosity length'' (e.g.\ Owocki, Gayley \&
Shaviv 2004). Note that for optically thin clumps the smooth-wind
opacity $\kappa_{\rm smooth}$ is recovered, while in the limit of large
$\tau_{\rm clump}$ the effective opacity becomes $\kappa_{\rm eff} =
h^{-1}$.

In the case of line formation we have to address the issue of Doppler
shifts. In our one-component model {\em the clumps are the wind}. Hence
clumps as such move with the wind velocity field $\varv(r)$. Due to the
supersonic expansion, rays of a given (observer's frame)
frequency can only interact with clumps near the  {\em surface of
constant radial velocity} (CRVS). The line opacity of all other clumps is
Doppler-shifted out of the resonance. Hence the porosity effect for
lines is very pronounced, as illustrated in Fig.\,\ref{fig:clumps}. 

An important issue is the Doppler broadening {\em within} a clump, which
widens the CRVS to a resonance {\em zone}. We describe the clump opacity
with a gaussian distribution corresponding to a Doppler-broadening
velocity $\varv_{\rm D}$ , which may schematically account for the velocity
distribution inside the clump caused by stochastic motion (thermal,
microturbulent), as well as by velocity gradients. Note that the degree
of porosity for line radiation depends on this parameter: for smaller
$\varv_{\rm D}$, the opacity profile of a clump peaks to a higher
maximum, resulting in a smaller effective opacity of the atmosphere.

The 1D hydrodynamic calculations shown in Fig.\,\ref{fig:shocks}
predict that within the dense shells the radial velocity gradient is
negative, and span typically over a velocity range of 100\,km/s or less.
When the motion is strictly radial, the transverse velocity gradient is
positive. Assuming that the clumps cover a cone angle of a few degrees
(seen from the center of the star), a wind speed of 
$\approx$2000\,km/s leads to a transverse expansion of a clump with 
less than 100\,km/s.  

Empirically it is known that the narrowest spectral features from
stellar winds, like the DACs (discrete absorption components), still
have typical widths corresponding to 50 ... 100\,km/s. Therefore values
of $\varv_{\rm D}$ in this range seem to be a reasonable choice.

A further important parameter is of course the clump size, which is 
related to the mean separation of the clumps, and to their total number
at one instant of time. Several observational and theoretical
arguments help to restrict this parameter, as discussed in Oskinova et
al.\ (2007). There we implemented the {\em macroclumping} formalism only
in the ``formal integral'' of the PoWR code, thus making the first
approximation that the non-LTE source function is not affected.

A motivation of this study was the so-called P\,{\sc v} problem. 
Bouret et al.\ (2005) and Fullerton et al.\ (2006) had pointed out that 
in O star spectra the P\,{\sc v} resonance line in the extreme UV 
is observed much weaker than predicted. They suggest to reduce
drastically the adopted mass-loss rates, and compensate for this by 
higher density-enhancement factors $D$ in the microclumping treatment. 
Resonance lines depend, like the electron-scattering wings discussed in 
Sect.\,\ref{sect:microclumping}, only linearly on density. 

\begin{figure}
\includegraphics[width=\textwidth]{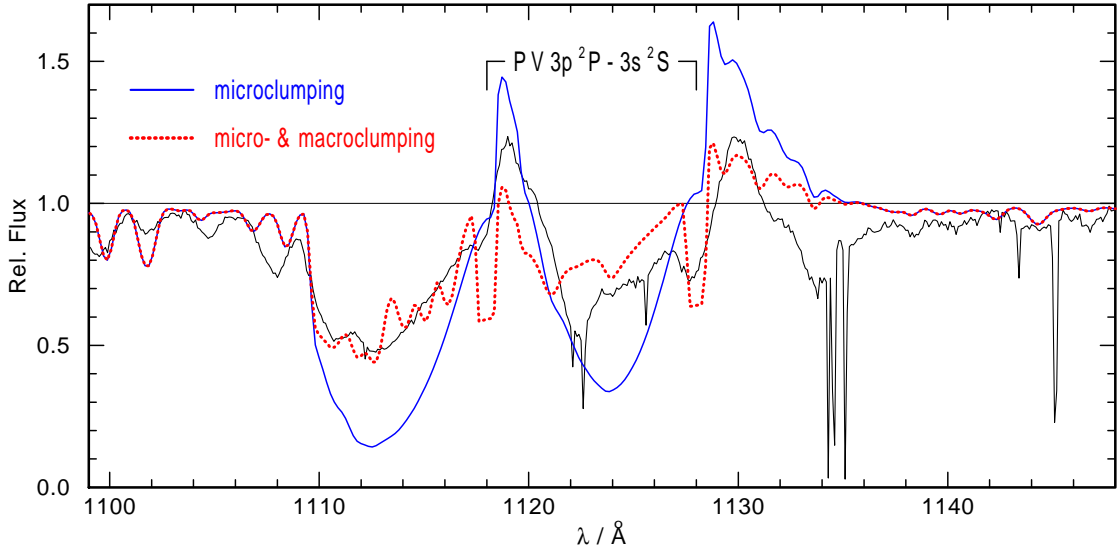}
\caption{Effect of macroclumping on the P\,{\sc v} resonance doublet at
1118/1128\,\AA\ for an O star. The spectrum of $\zeta$\,Pup as observed
by {\sc copernicus} is shown for comparison (black, ragged line). We
adopt a microturbulence velocity of 50\,km/s to account for the velocity
dispersion inside the clumps. Numerous weak spectral features in this
range are due to iron. The usual modeling yields much too strong
P\,Cygni features (blue, continuous line). With our macroclumping
formalism, the line features are reduced to the observed strength (red,
dotted curve). From Oskinova et al.\ (2007)}
\label{fig:zpup-pv}
\end{figure}

In Oskinova et al.\ (2007) we have shown that the macroclumping formalism 
can explain the weakness of the P\,{\sc v} resonance line without a 
further reduction of the mass-loss rate (see Fig.\,\ref{fig:zpup-pv}). 

%-----------------------------------------------------
%-----------------------------------------------------

As an unwanted side effect, macroclumping slightly de-saturates the black
absorptions of strong UV resonance lines like the doublets from C\,{\sc
iv} and N\,{\sc v}. However, according to Zsarg\'o et al.\ (2008) these
lines are anyhow not mainly formed in the dense clumps, but in the
diluted interclump medium which is neglected in our modeling.

%%%%%%%%%%%%%%%%%%%%%%%%%%%%%%%%%%%%%%%%%%%%%%%%
%% BACKMATTER
%%%%%%%%%%%%%%%%%%%%%%%%%%%%%%%%%%%%%%%%%%%%%%%%

%\begin{theacknowledgments}
%...
%\end{theacknowledgments}


\bibliographystyle{aipprocl} % if natbib is missing

%%%%%%%%%%%%%%%%%%%%%%%%%%%%%%%%%%%%%%%%%%%
%% The following lines show an example how to produce a bibliography
%% without the help of the BibTeX program. This could be used instead
%% of the above.
%%%%%%%%%%%%%%%%%%%%%%%%%%%%%%%%%%%%%%%%%%%

%\begin{thebibliography}{9}

\end{document}